# Sequential multiple importance sampling for high-dimensional Bayesian inference

Binbin Li[a,b], Xiao He[a,b], Zihan Liao[a,b,*]

[a] State Key Laboratory of Biobased Transportation Fuel Technology, ZJU-UIUC Institute, Zhejiang University, Haining, Zhejiang, China

[b] College of Civil Engineering and Architecture, Zhejiang University, Hangzhou, Zhejiang China

[*] Corresponding author; E-mail address: zihanliao@intl.zju.edu.cn

## Abstract

This paper introduces a sequential multiple importance sampling (SeMIS) algorithm for high-dimensional Bayesian inference. The method estimates Bayesian evidence using all generated samples from each proposal distribution while obtaining posterior samples through an importance-resampling scheme. A key innovation of SeMIS is the use of a softly truncated prior distribution as the intermediate proposal, providing a new way bridging prior and posterior distributions. By enabling samples from high-likelihood regions to traverse low-probability zones, SeMIS enhances mode mixing in challenging inference problems. Comparative evaluations against subset simulation (SuS) and adaptive Bayesian updating with structural reliability methods (aBUS) demonstrate that SeMIS achieves superior performance in evidence estimation (lower bias and variance) and posterior sampling (higher effective sample sizes and closer approximation to the true posterior), particularly for multimodal distributions. The efficacy of SeMIS is further validated in a high-dimensional finite element model updating application, where it successfully localizes structural damages by quantifying stiffness loss. The proposed algorithm not only advances Bayesian computation for complex posterior distributions but also provides a robust tool for uncertainty quantification in civil engineering systems, offering new possibilities for probabilistic structural health monitoring.

# 1. Introduction

Bayesian inference has been widely applied in civil engineering to update parameter $\boldsymbol{\theta} = [\theta_1, \theta_2, \cdots, \theta_n]^\mathrm{T} \in \mathbb{R}^n$ based on observed data $\boldsymbol{D}$ [1–3]. These parameters are typically related to structural characteristics like material properties, while the data often consist of measurements such as structural dynamic or static responses. In the Bayesian inference, the unknown parameter $\boldsymbol{\theta}$ is treated as a random vector with a probability density function (PDF) $\pi(\boldsymbol{\theta})$. The engineering knowledge is often encoded into $\pi(\boldsymbol{\theta})$ as a soft constraint to guide the inference. Since $\pi(\boldsymbol{\theta})$ does not incorporate the information from measured data, it is called the "prior" distribution. The data information is represented by the likelihood function $L(\boldsymbol{\theta}; \boldsymbol{D})$, which is understood as the likelihood of observing data $\boldsymbol{D}$ given parameter $\boldsymbol{\theta}$. Combining the information contained in engineering knowledge and data, one obtains the "posterior" PDF of unknown parameter $\boldsymbol{\theta}$ following the Bayes' theorem

$$p(\boldsymbol{\theta}|\boldsymbol{D}) = \frac{1}{z} L(\boldsymbol{\theta}; \boldsymbol{D}) \pi(\boldsymbol{\theta}) \tag{1}$$

where $z = \int L(\boldsymbol{\theta}; \boldsymbol{D}) \pi(\boldsymbol{\theta}) \mathrm{d}\boldsymbol{\theta}$ is a normalization constant so that $p(\boldsymbol{\theta}|\boldsymbol{D})$ is a valid PDF. It is also called as the evidence, quantifying how likely the observed data are under the current model. Thus, it can be used for model selection. The posterior PDF $p(\boldsymbol{\theta}|\boldsymbol{D})$ quantifies the remaining uncertainty associated with the parameter $\boldsymbol{\theta}$. The Bayesian inference provides a systematic framework to combine data and engineering knowledge for a continuous update of the belief on the parameter $\boldsymbol{\theta}$.

It is generally difficult to compute the posterior distribution $p(\boldsymbol{\theta}|\boldsymbol{D})$ and the evidence $z$, because the likelihood function is usually a highly nonlinear function of the parameter $\boldsymbol{\theta}$. Direct approximation methods, e.g., Laplace approximation [4] and kernel density estimation [5], have been developed to fit the posterior distribution. Since these methods assume relatively simple model characteristics, e.g., unimodal and approximately normal, they do not perform well in dealing with complex structural models or limited data, where the posterior distribution may exhibit multiple modes or heavy tails. Monte Carlo simulation is another frequently employed method for posterior approximation using random samples. Since weaker assumptions are made in Monte Carlo simulation, it is more suitable to address complex posterior distributions, which is the focus of this paper.

Since the direct sampling from the posterior distribution is generally hard, annealing-based strategies have been proposed in Bayesian inference to construct intermediate distributions that allow samples to transit smoothly from the prior to the posterior distribution. Examples of such methods include path sampling [6], stepping-stone sampling [7], power posteriors [8] and transitional Markov Chain Monte Carlo (MCMC) [9]. This smooth transformation facilitates better mixing of samples, enhancing the sampling ergodicity. However, constructing these intermediate distributions often requires careful tuning, and suboptimal settings can lead to weight degeneration and biased estimates [10].

Vertical likelihood-based approach provides an alternative method in defining the intermediate distributions, as seen in the nested sampling [11] (as well as its variants MULTINEST [12] and POLYCHORD [13]) and the subset simulation (SuS) [14]. These methods iteratively truncate the sampling region based on adaptively defined likelihood thresholds, effectively concentrating computational effort in areas of high likelihood [15]. Although this strategy efficiently focuses on important samples, the hard truncation can easily create isolated regions in the parameter space, leading to inadequate sampling across different posterior modes. Bayesian updating with structural reliability methods (BUS) [16,17] transforms the Bayesian inference into a reliability estimation problem. Although truncation still occurs, it is applied within an augmented space encompassing both the likelihood function and the likelihood coordinates, effectively maintaining connectivity in the parameter space. This method offers efficiency comparable to hard truncation algorithms while keeping the sampling region connected [14].

Inspired by the soft truncation in BUS and the framework of SuS, this paper proposes a sequential multiple importance sampling (SeMIS) algorithm for Bayesian inference. It aims at combining the advantages of sample mixing in annealing-based approaches with the efficiency of concentrating samples in high-likelihood areas, as in the vertical likelihood-based approaches. This paper is organized as follows: Section 2 reviews the multiple importance sampling (MIS) framework for Bayesian inference;

Section 3 provides a detailed introduction of our SeMIS algorithm; Section 4 demonstrates the performance of the proposed algorithm in Bayesian inference for three benchmark examples and damage detection of a benchmark building model. Final conclusions are provided in Section 5.

## 2. Bayesian inference with multiple importance sampling

In this section, we will first review the MIS framework for Bayesian evidence evaluation and posterior sampling [18–21], and then re-interpret the SuS algorithm [14] as a special case of MIS. This provides the mathematical foundation for the development of SeMIS algorithm.

### 2.1 Multiple importance sampling

Following the integral of evidence in Eqn. (1), we first modify it by introducing a weighting scheme in terms of $\alpha_i(\boldsymbol{\theta})$ $(i = 0, \dots, I-1)$ to obtain

$$z = \int \sum_{i=0}^{I-1} \alpha_i(\boldsymbol{\theta}) L(\boldsymbol{\theta})\pi(\boldsymbol{\theta})\, d\boldsymbol{\theta} = \sum_{i=0}^{I-1} \int \alpha_i(\boldsymbol{\theta}) L(\boldsymbol{\theta})\pi(\boldsymbol{\theta})\, d\boldsymbol{\theta} \tag{2}$$

where $\alpha_i(\boldsymbol{\theta})$ is called the proposal weight, satisfying $\sum_{i=0}^{I-1} \alpha_i(\boldsymbol{\theta}) = 1$ when $L(\boldsymbol{\theta})\pi(\boldsymbol{\theta}) \neq 0$. For each $\alpha_i(\boldsymbol{\theta})$, we then introduce a proposal distribution $q_i(\boldsymbol{\theta})$, resulting in

$$z = \sum_{i=0}^{I-1} \int \alpha_i(\boldsymbol{\theta}) \frac{L(\boldsymbol{\theta})\pi(\boldsymbol{\theta})}{q_i(\boldsymbol{\theta})} q_i(\boldsymbol{\theta})\, d\boldsymbol{\theta} = \sum_{i=0}^{I-1} \underbrace{\int \alpha_i(\boldsymbol{\theta}) w_i(\boldsymbol{\theta}) q_i(\boldsymbol{\theta})\, d\boldsymbol{\theta}}_{z_i} \tag{3}$$

where $w_i(\boldsymbol{\theta}) = \pi(\boldsymbol{\theta})L(\boldsymbol{\theta})/q_i(\boldsymbol{\theta})$ is an unnormalized posterior weight function, and $z_i$ represents the contribution to the evidence from the $i$-th proposal distribution. Note that the validation of Eqn. (3) requires $\alpha_i(\boldsymbol{\theta}) = 0$ when $q_i(\boldsymbol{\theta}) = 0$.

To estimate the evidence contribution $z_i$, one can generate a random sample of size $N_i$ from the proposal distribution $q_i(\boldsymbol{\theta})$, denoted as $\{\boldsymbol{\Theta}_{i,1}, \boldsymbol{\Theta}_{i,2}, \dots, \boldsymbol{\Theta}_{i,N_i}\}$, to construct the following estimator

$$\hat{Z}_i = \frac{1}{N_i} \sum_{k=1}^{N_i} \alpha_i(\boldsymbol{\Theta}_{i,k}) w_i(\boldsymbol{\Theta}_{i,k}) \tag{4}$$

Since $\hat{Z}_i$ is an unbiased estimator of $z_i$, the overall MIS estimator $\hat{Z} = \sum_{i=0}^{I-1} \hat{Z}_i$ is also unbiased. The estimation variance is given by

$$\text{Var}\{\hat{Z}\} = \sum_{i=0}^{I-1} \frac{1}{N_i} \underbrace{\int [\alpha_i(\boldsymbol{\theta})w_i(\boldsymbol{\theta}) - z_i]^2 q_i(\boldsymbol{\theta})d\boldsymbol{\theta}}_{\sigma_i^2} \tag{5}$$

Assuming $\boldsymbol{\Theta}_{i,k}$'s are generated independently from $q_i(\boldsymbol{\theta})$, the variance $\sigma_i^2$ can be estimated by

$$\hat{\Sigma}_i = \frac{1}{N_i}\sum_{k=1}^{N_i}[\alpha_i(\boldsymbol{\Theta}_{i,k})w_i(\boldsymbol{\Theta}_{i,k}) - \hat{Z}_i]^2 \tag{6}$$

By properly selecting the proposal weights $\alpha_i(\boldsymbol{\theta})$, proposal distributions $q_i(\boldsymbol{\theta})$ and the sample size $N_i$, the MIS estimator can reduce the variance of crude Monte Carlo and the conventional importance sampling (a special case of MIS with $I = 1$).

To obtain posterior samples, we first re-write the evidence estimator as

$$\hat{Z} = \frac{1}{N_t}\sum_{i=1}^{I-1}\sum_{k=1}^{N_i} w_{i,k} \tag{7}$$

where $N_t = \sum_{i=0}^{I-1} N_i$ is the total sample size, and $w_{i,k}$ is given by

$$w_{i,k} = \frac{N_t}{N_i}w_i(\boldsymbol{\Theta}_{i,k})\alpha_i(\boldsymbol{\Theta}_{i,k}) \tag{8}$$

Equation (7) can be interpreted as an importance sampling estimator for the evidence, where the proposal distribution corresponds to a mixture distribution $\sum_{i=0}^{I-1} N_i q_i(\boldsymbol{\theta})/N_t$ from which the overall sample $\{\boldsymbol{\Theta}_{i,k}, k = 1,2,\ldots,N_i, i = 0,1,2,\ldots,I-1\}$ is drawn. In this context, $w_{i,k}$ is proportion to the ratio of posterior PDF to the proposal PDF. This implies that $w_{i,k}$ corresponds to the unnormalized posterior weight of sample $\boldsymbol{\Theta}_{i,k}$. Consequently, posterior samples can be obtained through an importance resampling step based on weights $w_{i,k}$.

The number of posterior samples that can be generated is related to the effective sample size (ESS), denoted $N_{ess}$, which can be understood as the maximum number of independent posterior samples [22]. If the samples $\boldsymbol{\Theta}_{i,k}$ were generated independently from $q_i(\boldsymbol{\theta})$ in MIS, $N_{ess}$ can be approximate as [10]

$$N_{ess} \approx \frac{(\sum_i \sum_k w_{i,k})^2}{\sum_i \sum_k w_{i,k}^2} \tag{9}$$

When samples $\boldsymbol{\theta}_{i,k}$'s are statistically dependent, e.g., when they are generated from an MCMC sampler, the $N_{ess}$ in Eqn. (9) gives an upper bound on the number of independent posterior samples. This will be used as a performance index to compare various algorithms for Bayesian inference.

**2.2 Re-interpretation of SuS**

SuS is a recently developed algorithm for Bayesian inference [14]. It is based on a new interpretation of the evidence in Eqn. (1) from the perspective of reliability estimation:

$$\begin{aligned} z &= \int_0^{L_{\sup}} \underbrace{\left[\int_\Omega \mathbb{1}[L(\boldsymbol{\theta}) > l]\pi(\boldsymbol{\theta})\mathrm{d}\boldsymbol{\theta}\right]}_{p_f(l)} \mathrm{d}l \\ &= \sum_{i=0}^{I-1} \wp_i \int \min\{L(\boldsymbol{\theta}) - l_i, l_{i+1} - l_i\} q_i^{(\mathrm{SuS})}(\boldsymbol{\theta})\mathrm{d}\boldsymbol{\theta} \end{aligned} \tag{10}$$

where $p_f(l)$ represents a failure probability with threshold $l$ and limit state function $L(\boldsymbol{\theta})$ (i.e., likelihood in Bayesian inference). Here, "$\mathbb{1}[\cdot]$" denotes the indicator function, which returns one if the inequality is satisfied and zero otherwise. The maximum value of $L(\boldsymbol{\theta})$, denoted $L_{\sup}$, is generally unknown but plays no important role in the practical implementation, because $p_f(l)$ quickly decreases to zero as the threshold $l$ increases. The second equation is derived by segmenting the integration domain as $[0, L_{\sup}] = [l_0, l_1] \cup (l_1, l_2] \cup \ldots \cup (l_{I-1}, l_I]$ with $l_0 = 0$ and $l_I = L_{\sup}$. The proposal distribution $q_i^{(\mathrm{SuS})}(\boldsymbol{\theta})$ is defined as

$$q_i^{(\mathrm{SuS})}(\boldsymbol{\theta}; l_i) = \wp_i^{-1}\pi(\boldsymbol{\theta})\mathbb{1}[L(\boldsymbol{\theta}) > l_i] \tag{11}$$

where $\wp_i$ works as a normalization factor so that $q_i^{(\mathrm{SuS})}(\boldsymbol{\theta})$ is a valid PDF. Since $l_0 = 0$, one has the first proposal distribution $q_0^{(\mathrm{SuS})}(\boldsymbol{\theta}) = \pi(\boldsymbol{\theta})$, i.e., the prior distribution. In the SuS algorithm, samples from $q_i^{(\mathrm{SuS})}(\boldsymbol{\theta})$ are generated following a parallel MCMC scheme, and $l_{i+1}$ is set as the $(1 - p_c)$ quantile of the likelihood of generated samples, where $p_c$ is called the level probability (with a default value of 0.1).

Comparing Eqn. (10) with Eqn. (3), one can find that the SuS algorithm is effectively a special case of MIS with the proposal distribution $q_i^{(\text{SuS})}(\boldsymbol{\theta})$ in Eqn. (11) and the proposal weight

$$\alpha_i^{(\text{SuS})}(\theta) = \frac{\min\{L(\boldsymbol{\theta}) - l_i, l_{i+1} - l_i\}\mathbb{1}[L(\boldsymbol{\theta}) > l_i]}{L(\boldsymbol{\theta})} \tag{12}$$

To verify the above statement, one should have $\sum_{i=0}^{I-1} \alpha_i^{(\text{SuS})}(\theta) = 1$, or equivalently

$$\sum_{i=0}^{I-1} \min\{L(\boldsymbol{\theta}) - l_i, l_{i+1} - l_i\}\mathbb{1}[L(\boldsymbol{\theta}) > l_i] = L(\boldsymbol{\theta}) \tag{13}$$

Without loss of generality, considering $l_j < L(\boldsymbol{\theta}) \leq l_{j+1}$, the left-hand side (LHS) of Eqn. (13) reads

$$\sum_{i=0}^{I-1} \min\{L(\boldsymbol{\theta}) - l_i, l_{i+1} - l_i\}\mathbb{1}[L(\boldsymbol{\theta}) > l_i] = \sum_{i=0}^{j} \min\{L(\boldsymbol{\theta}) - l_i, l_{i+1} - l_i\}\mathbb{1}[L(\boldsymbol{\theta}) > l_i]$$
$$= \sum_{i=0}^{j-1}(l_{i+1} - l_i) + [L(\boldsymbol{\theta}) - l_j] = L(\boldsymbol{\theta}) \tag{14}$$

where we have reduced the summation limit from $I - 1$ to $j$ in the first equation, and evaluated each term in the second equation to finally yield $L(\boldsymbol{\theta})$ with $l_0 = 0$. Therefore, the LHS of Eqn. (13) effectively represents a segmented decomposition of $L(\boldsymbol{\theta})$, confirming that the SuS algorithm is a specific form of MIS.

Although the SuS algorithm works well in many examples, it is not optimal from the perspective of MIS. The chosen proposal weight $\alpha_i^{(\text{SuS})}(\boldsymbol{\theta})$ does not minimize the estimation variance, i.e., not efficient. The proposal distribution $q_i^{(\text{SuS})}(\boldsymbol{\theta})$ is in a form of truncated prior distribution, typically creating isolated domains in the parameter space. Since it is difficult for a MCMC sampler to frequently transition between different modes, it poses challenges for ergodicity in sampling from $q_i^{(\text{SuS})}(\boldsymbol{\theta})$ using MCMC samplers. With the MIS interpretation, we will address these limitations of SuS in the next section.

## 3. Sequential multiple importance sampling

Inspired by SuS and BUS algorithms, we propose the SeMIS algorithm in this section, incorporating softly truncated prior for proposal distributions and the balance heuristic for proposal weights [23,24]. Section 3.1 introduces the general setting, and outlines the main procedure. Section 3.2 provides the key implementation details.

**3.1 Main algorithm**

The proposed SeMIS algorithm is typical MIS algorithm with the following proposal distribution

$$q_i(\boldsymbol{\theta}; \lambda_i) = \wp_i^{-1}\pi(\boldsymbol{\theta})L_i(\boldsymbol{\theta}; \lambda_i) \tag{15}$$

where $\wp_i$ is a normalization constant ensuring $q_i(\boldsymbol{\theta}; \lambda_i)$ is a valid PDF. The term $L_i(\boldsymbol{\theta}; \lambda_i)$ is a compound function involving with the likelihood function as

$$L_i(\boldsymbol{\theta}; \lambda_i) = \min\left[\frac{L(\boldsymbol{\theta})}{\lambda_i L_{\sup}}, 1\right] \tag{16}$$

where $L_{\sup}$ denotes the supreme value of the likelihood function $L(\boldsymbol{\theta})$, and $\lambda_i$ is a hyperparameter value ranging from zero to one. Note that the proposal distribution $q_i(\boldsymbol{\theta}; \lambda_i)$ corresponds to the prior $\pi(\boldsymbol{\theta})$ if $\lambda_i = 0$, while it becomes the posterior distribution $p(\boldsymbol{\theta}|\boldsymbol{D})$ if $\lambda_i = 1$. That is, the proposal distribution $q_i(\boldsymbol{\theta}; \lambda_i)$, in fact, defines a combined distribution of prior and posterior. In SeMIS, we require the hyperparameter $\lambda_i$ to monotonically increase from zero to one, thus creating a path from the prior to the posterior, as illustrated in Fig. 1 a) for a bi-modal posterior distribution. Additionally, the increasing $\lambda_i$ ensures that samples with higher likelihood values play an increasingly important role, which is similar to the SuS algorithm.

For comparison, we also draw in Fig. 1 b) the proposal distributions $q_i^{(\text{SuS})}(\boldsymbol{\theta}; l_i)$ in SuS by intentionally setting $l_i = \lambda_i L_{\sup}$. It is seen that $q_i^{(\text{SuS})}(\boldsymbol{\theta}; l_i)$ corresponds to a hard truncated prior (uniform distribution in this case), which quickly becomes isolated in regions defined by $L(\boldsymbol{\theta}) > l_i$. Note that $q_i^{(\text{SuS})}(\boldsymbol{\theta}; l_i)$ does not include the posterior distribution, which is illustrated here only for an easier comparison. As a counterpart, $q_i(\boldsymbol{\theta}; \lambda_i)$ defined in Eqn. (15) is a mixture of prior and posterior distributions, and it has larger tails in low values of $L(\boldsymbol{\theta})$, improving the connectiveness of different modes. This behavior allows for the transition of samples from high likelihood regions to low likelihood regions, facilitating mixing between different modes in MCMC sampling and enhancing the ergodicity when a limited number of samples are used.

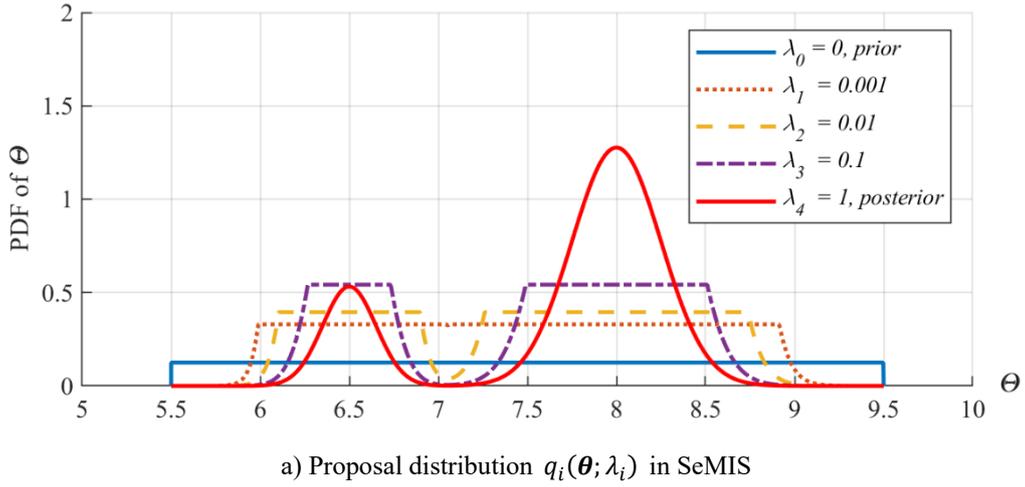

a) Proposal distribution $q_i(\boldsymbol{\theta}; \lambda_i)$ in SeMIS

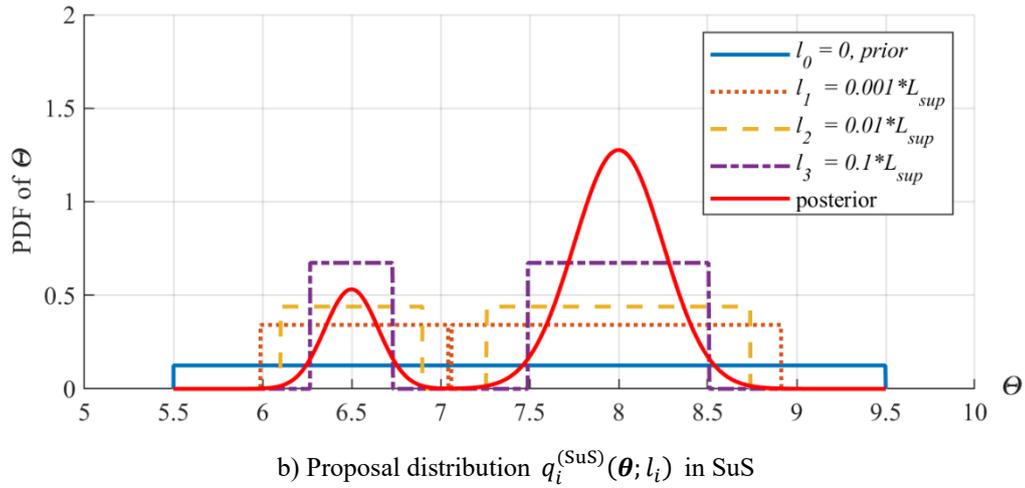

b) Proposal distribution $q_i^{(\text{SuS})}(\boldsymbol{\theta}; l_i)$ in SuS

Figure 1. Evolution of proposal distributions

Note: prior PDF Unif(5.5,9.5), posterior PDF $0.8N(8, 0.25^2) + 0.2N(6.5, 0.15^2)$

Within the MIS framework, a classic approach known as the balance heuristic [23,24] is adopted to construct the proposal weights $\alpha_i(\boldsymbol{\theta})$ as

$$\alpha_i(\boldsymbol{\theta}) = \frac{N_i q_i(\boldsymbol{\theta}; \lambda_i)}{\sum_{j=0}^{I-1} N_j q_j(\boldsymbol{\theta}; \lambda_j)} \tag{17}$$

where $N_i$ is the sample size generated from $q_i(\boldsymbol{\theta}; \lambda_i)$. By substituting Eqn. (17) into Eqn. (4), we can derive the following evidence estimator

$$\hat{Z}_{\mathrm{MIS}} = \sum_{i=0}^{I-1} \frac{1}{N_i} \sum_{k=1}^{N_i} \frac{L(\boldsymbol{\Theta}_{i,k})\pi(\boldsymbol{\Theta}_{i,k})}{q_i(\boldsymbol{\Theta}_{i,k}; \lambda_i)} \frac{N_i q_i(\boldsymbol{\Theta}_{i,k}; \lambda_i)}{\sum_{j=0}^{I-1} N_j q_j(\boldsymbol{\Theta}_{i,k}; \lambda_j)} = \sum_{i=0}^{I-1} \sum_{k=1}^{N_i} \frac{L(\boldsymbol{\Theta}_{i,k})\pi(\boldsymbol{\Theta}_{i,k})}{\sum_{j=0}^{I-1} N_j q_j(\boldsymbol{\Theta}_{i,k}; \lambda_j)} \quad (18)$$

which utilizes all samples generated from each proposal distribution. Note that the proposal distribution $q_i(\boldsymbol{\theta}; \lambda_i)$ is sequentially and adaptively determined as detailed in Section 3.2.1. Therefore, the overall algorithm formulates a sequential version of MIS, explaining its name SeMIS. The main procedures of SeMIS algorithm are summarized in Algorithm 1 in terms of a pseudocode.

Algorithm 1: SeMIS algorithm for Bayesian inference

---

1. Initialization

    Set the expected acceptance rate $p$; % default 0.1, used to determine the hyperparameter $\lambda_i$

    Set the target sample size $N$ for each proposal distribution; % default 1000

2. Iteration 0

(1) Generate $N_0 = N$ independent random samples $\{\boldsymbol{\theta}_{0,k}: k = 1,2,\dots,N_0\}$ from the prior distribution $\pi(\boldsymbol{\theta})$, and calculate the corresponding likelihood $y_{0,k} = L(\boldsymbol{\theta}_{0,k})$.

(2) Find the hyperparameter $\lambda_1$ and determine the proposal PDF $q_1(\boldsymbol{\theta}; \lambda_1)$; % see Section 3.2.1

3. **For** Iteration $i = 1,2,\dots,I$    % $I$ is the predefined maximum number of proposal distributions

(1) Calculate the acceptance rate corresponding to each sample $\beta_{i,k} = q_i(\boldsymbol{\theta}_{i,k}; \lambda_i)/cq_{i-1}(\boldsymbol{\theta}_{i,k}; \lambda_{i-1})$, and select the seed samples $\{\boldsymbol{\theta}_{i,j,0}, j = 1,2,\dots,N_i^{(c)}\}$. % see Section 3.2.2

(2) For each seed sample $\boldsymbol{\theta}_{i,j,0}$, generate a Markov chain of length $N_i^{(s)}$ using a parallel MCMC algorithm to obtain samples $\{\boldsymbol{\theta}_{i,k}, k = 1,2,\dots,N_i^{(c)}N_i^{(s)}\}$, and calculate the corresponding likelihood $y_{i,k} = L(\boldsymbol{\theta}_{i,k})$. % see Section 3.2.2

(3) Update hyperparameters $\lambda_{i+1}$ and determine $q_{i+1}(\boldsymbol{\theta}; \lambda_{i+1})$ based on the expected acceptance rate $p$. % see Section 3.2.1

(4) **If** convergence criterion is met, **STOP**; **Endif**    % see Section 3.2.3

    **Endfor**

4. Postprocessing

    Estimate the evidence $\hat{Z}_{\mathrm{MIS}}$ based on Eqn. (18);

    Regenerate posterior samples $\{\boldsymbol{\theta}_k: k = 1,2,\dots,N_{ess}\}$ according to the posterior weight $w_{i,k}$    % $N_{ess}$ denotes the effective sample size as shown in Eqn. (9).

---

First, two parameters are needed for initialization: number of samples $N$ and expected acceptance rate $p$. Here, $N$ is only the targeted number of samples from each proposal distribution, but the practical number of samples may vary around $N$ due to the adopted parallel MCMC algorithm, as describe later. The expected acceptance rate $p$, similar to the level probability in SuS, defines the relation between two successive proposal distributions as

$$p = \mathbb{E}_{q_{i-1}(\boldsymbol{\theta};\lambda_{i-1})}[\beta_i(\boldsymbol{\theta})] = \int q_{i-1}(\boldsymbol{\theta};\lambda_{i-1}) \underbrace{\frac{q_i(\boldsymbol{\theta};\lambda_i)}{cq_{i-1}(\boldsymbol{\theta};\lambda_{i-1})}}_{\beta_i(\boldsymbol{\theta})} \mathrm{d}\boldsymbol{\theta} \qquad (19)$$

where $c = \max[q_i(\boldsymbol{\theta};\lambda_i)/q_{i-1}(\boldsymbol{\theta};\lambda_{i-1})] = \wp_{i-1}/\wp_i$ is a constant that ensures the maximum value of $\beta_i(\boldsymbol{\theta})$ is 1. This is adopted from the acceptance-rejection sampling (ARS) principle [25–27] to ensure two successive proposal distributions $q_i(\boldsymbol{\theta};\lambda_i)$ and $q_{i-1}(\boldsymbol{\theta};\lambda_{i-1})$ are not too different from each other. It aims to create a smooth transition using samples generated from $q_{i-1}(\boldsymbol{\theta};\lambda_{i-1})$ as seeds to sample from $q_i(\boldsymbol{\theta};\lambda_i)$. The value of $p = 0.1$ is selected in this paper for all considered examples.

The SeMIS algorithm then starts from a random sampling from the prior PDF $\pi(\boldsymbol{\theta})$, from which the first proposal distribution $q_1(\boldsymbol{\theta};\lambda_1)$ is adaptively determined. The parallel MCMC is adopted to sample from $q_1(\boldsymbol{\theta};\lambda_1)$ and determine the next proposal distribution $q_2(\boldsymbol{\theta};\lambda_2)$. The above procedures are iterated until convergence is achieved. Detailed explanation of above procedures is provided in the following subsections.

### 3.2 Key elements

The content in Section 3.1 outlines the procedure of SeMIS, with some details omitted in the pseudocode. This section will explore in depth the key elements of the proposed algorithm, including proposal distribution determination, parallel MCMC sampling and termination conditions.

#### 3.2.1 Adaptive determination of proposal distribution

One main difference of the proposed algorithm from the conventional MIS is the adaptive determination of proposal distribution, i.e., sequentially find $q_i(\boldsymbol{\theta};\lambda_i)$ given $q_{i-1}(\boldsymbol{\theta};\lambda_{i-1})$. It is determined to achieve a targeted expected acceptance rate $p$ as shown in Eqn. (19). To obtain such $\lambda_i$, we first simply the expression of acceptance rate $\beta_i(\boldsymbol{\theta})$ by substituting Eqn. (15) into Eqn. (19) as

$$\beta_i(\boldsymbol{\theta}) = \frac{\wp_{i-1}}{c\wp_i} \frac{\min\left[L(\boldsymbol{\theta})/(\lambda_{i-1}L_{\sup}),1\right]}{\min\left[L(\boldsymbol{\theta})/(\lambda_i L_{\sup}),1\right]} \tag{20}$$

and the constant

$$c = \max\left[\frac{q_i(\boldsymbol{\theta};\lambda_i)}{q_{i-1}(\boldsymbol{\theta};\lambda_{i-1})}\right] = \frac{\wp_{i-1}}{\wp_i}\max\left\{\frac{\min\left[L(\boldsymbol{\theta})/(\lambda_i L_{\sup}),1\right]}{\min\left[L(\boldsymbol{\theta})/(\lambda_{i-1}L_{\sup}),1\right]}\right\} = \frac{\wp_{i-1}}{\wp_i} \tag{21}$$

because the maximum of $\min\left[L(\boldsymbol{\theta})/(\lambda_i L_{\sup}),1\right]/\min\left[L(\boldsymbol{\theta})/(\lambda_{i-1}L_{\sup}),1\right]$ is 1 due to $\lambda_i > \lambda_{i-1}$. Therefore, $\beta_i(\boldsymbol{\theta})$ can be finally simplified as

$$\beta_i(\boldsymbol{\theta}) = \frac{\min\left[L(\boldsymbol{\theta})/(\lambda_{i-1}L_{\sup}),1\right]}{\min\left[L(\boldsymbol{\theta})/(\lambda_i L_{\sup}),1\right]} \tag{22}$$

For the expectation of $\beta_i(\boldsymbol{\theta})$, since the proposal distribution $q_{i-1}(\boldsymbol{\theta};\lambda_{i-1})$ does not belong to any standard distribution, one cannot evaluate it analytically. Alternatively, the Monte Carlo approximation is used based on the random sample $\{\boldsymbol{\theta}_{i-1,k}, k = 1,\dots,N_{i-1}\}$, i.e., $\lambda_i$ satisfies $\sum_{k=1}^{N_{i-1}}\beta_i(\boldsymbol{\theta}_{i-1,k})/N_{i-1} = p$.

It should be noted that the proposal distribution $q_i(\boldsymbol{\theta};\lambda_i)$ also depends on $L_{\sup}$, i.e., the supreme value of the likelihood function $L(\boldsymbol{\theta})$, which is unknown a priori in most problems. In this paper, we consider an equivalent parameterization $\lambda_i L_{\sup} = r_i l_{0:i-1}^{\max}$ ($r_0 = 0$ and $l_{0:-1}^{\max} = 0$ for $i = 0$), where $l_{0:i-1}^{\max}$ represents the maximum value of the likelihood of all generated samples until to the $(i-1)$-th step and $r_i$ ($= \lambda_i L_{\sup}/l_{0:i-1}^{\max} \geq \lambda_i$) is a coefficient belonging to the interval $[0,1]$. Given a random sample $\{\boldsymbol{\theta}_{i-1,k}, k = 1,\dots,N_{i-1}\}$ drawn from $q_i(\boldsymbol{\theta};\lambda_i)$, one can evaluate $l_{0:i-1}^{\max} = \max\{l_{0:i-2}^{\max},\{y_{i-1,k}\}\}$ with $y_{i-1,k} = L(\boldsymbol{\theta}_{i-1,k})$. The problem of determining $\lambda_i$, thus, reduces to the determination of $r_i$, which satisfies $r_i > r_{i-1}l_{0:i-2}^{\max}/l_{0:i-1}^{\max}$ and $\sum_{k=1}^{N_{i-1}}\beta_i(\boldsymbol{\theta}_{i-1,k})/N_{i-1} = p$. The former condition ensures a corresponding monotonically increasing sequence of $\lambda_i$'s, as required in the SeMIS algorithm. Note that searching an appropriate value of $r_i$ does not involve with evaluating the likelihood function, so that the computational cost is not a concern, leaving the choice of optimization method flexible. In our application, we simply apply the MATLAB function "fminbnd" to solve the constrained minimization problem $\min_{r_{i-1}l_{0:i-2}^{\max}/l_{0:i-1}^{\max} \leq r_i \leq 1}\left|\sum_{k=1}^{N_{i-1}}\beta_i(\boldsymbol{\theta}_{i-1,k})/N_{i-1} - p\right|$. In the following, we keep the notation

of $q_i(\boldsymbol{\theta}; \lambda_i)$ for the proposal distribution, because it provides a better understanding of the nature of proposal distributions, but use $r_i$ and $l_{0:i-1}^{\max}$ in implementation.

**3.2.2 Parallel MCMC sampling**

In the SeMIS algorithm, it is crucial to efficiently generate random samples from the proposal distribution $q_i(\boldsymbol{\theta}; \lambda_i)$, particularly when the parameter space of $\boldsymbol{\theta}$ is high-dimensional. To address this, we use a parallel MCMC sampling approach that runs multiple short Markov chains. This method involves two key steps: seed selection and MCMC sampling, described below.

The seed selection follows an ARS [25–27] principle to take advantage of the existing random sample $\{\boldsymbol{\theta}_{i-1,k}, k = 1, \dots, N_{i-1}\}$ drawn from $q_{i-1}(\boldsymbol{\theta}; \lambda_{i-1})$. Each $\boldsymbol{\theta}_{i-1,k}$ is accepted as a seed with $\beta_i(\boldsymbol{\theta}_{i-1,k})$, calculated via Eqn. (22). Based on the ARS principle, the above screen process ensures seeds approximately follow the target distribution $q_i(\boldsymbol{\theta}; \lambda_i)$, significantly reducing the need for a traditional MCMC "burn-in" phase. Note that the resulted seeds $\{\boldsymbol{\theta}_{i,j,0}, j = 1,2, \dots, N_i^{(c)}\}$ are statistically correlated, which is different from the conventional ARS. Here, $N_i^{(c)}$ denotes the total number of accepted samples, which is expected to be around $pN$ following the initial setting.

To generate approximately $N$ new samples from the proposal distribution $q_i(\boldsymbol{\theta}; \lambda_i)$, we initiate $N_i^{(c)}$ parallel MCMC chains from selected seeds, with an identical length of $N_i^{(s)}$. To obtain an integer $N_i^{(s)}$ such that $N_i^{(c)} N_i^{(s)} \approx N$, we first compute candidate seed counts as $\text{round}([N/1, N/2, \dots, 1])$, where "round(·)" denotes the rounding function. The largest integer in this sequence that does not exceed $N_i^{(c)}$ is selected and assigned as the updated $N_i^{(c)}$. Next, $N_i^{(s)}$ is set to $\text{round}(N/N_i^{(c)})$. Finally, seeds are randomly re-selected with equal weighting to form the final seed set. This method ensures robustness and guarantees the desired sample size around $N$, regardless of the initial number of seeds.

For the MCMC sampling, we adopt the elliptical slice (ES) sampling algorithm [28], for its merit of no repeated samples (decreased statistical correlation) and no tuning parameters (ease in implementation). A comparison of ES sampling with adaptive conditional sampling and Hamiltonian Monte Carlo can be found in Ref. [29]. The ES sampling algorithm was developed for high-dimensional Bayesian inference with a normal prior. For this, one can apply the Rosenblatt transformation [30] or the marginal transformation based on the Nataf model [31] to define a transformation $\boldsymbol{\Theta} = T(\boldsymbol{U})$ such that $\boldsymbol{U} \sim \phi_n(\boldsymbol{u})$ while $\boldsymbol{\Theta} \sim \pi(\boldsymbol{\theta})$, which is typically done in reliability estimation. Here, $\phi_n(\boldsymbol{u})$ denotes an $n$-dimensional standard normal PDF. For completeness, the pseudocode of ES sampling is provided in Algorithm 2.

---

Algorithm 2:   Elliptical slice sampling

---

1. Initialization

   Given: The target distribution $q_i(\boldsymbol{\theta}; \lambda_i)$, the transformation $\boldsymbol{\Theta} = T(\boldsymbol{U})$ and inverse transformation $\boldsymbol{U} = T^{-1}(\boldsymbol{\Theta})$;

   Compute seeds in the standard normal space $\{\boldsymbol{u}_{i,j,0} = T^{-1}(\boldsymbol{\theta}_{i,j,0}), j = 1,2, \dots, N_i^{(c)}\}$;

2. **For** Iteration $j = 1,2, \dots, N_i^{(c)}$    % $N_i^{(c)}$ denotes # MCMC chains

   **For** Iteration $t = 0,1, \dots, N_i^{(s)} - 1$    % $N_i^{(s)}$ denotes # samples in each MCMC chain

   % Implementation in log-likelihood for numerical stability

   (1) Log-likelihood threshold: Draw $\gamma \sim \text{Unif}[0,1]$
   $$\ln y = \min\left[\ln L(T(\boldsymbol{u}_{i,j,t})) - \ln l_{0:i-1}^{\max} - \ln r_i, 0\right] + \ln\left|\frac{\partial T(\boldsymbol{u}_{i,j,t})}{\partial \boldsymbol{u}}\right| + \ln \gamma$$
   % The 2nd term is log determinant of Jacobian matrix evaluated at $\boldsymbol{u} = \boldsymbol{u}_{i,j,t}$

   (2) Draw a random angle $\alpha \sim \text{Unif}[0, 2\pi]$, and set $\alpha_{\min} = \alpha - 2\pi$ and $\alpha_{\max} = \alpha$;
   (3) Draw a random sample $\boldsymbol{v} \sim \phi_n(\boldsymbol{v})$;
   (4) Compute the candidate sample
   $$\boldsymbol{\xi} = \boldsymbol{u}_{i,j,t} \cos \alpha + \boldsymbol{v} \sin \alpha$$
   and log-likelihood: $\ln l = \min\left[\ln L(T(\boldsymbol{\xi})) - \ln l_{0:i-1}^{\max} - \ln r_i, 0\right] + \ln\left|\frac{\partial T(\boldsymbol{\xi})}{\partial \boldsymbol{u}}\right|$;

   (5) **If** $\ln l > \ln y$
       Accept: **return** $\boldsymbol{u}_{i,j,t+1} = \boldsymbol{\xi}$ and $\boldsymbol{\theta}_{i,j,t+1} = T(\boldsymbol{\xi})$;
   **Else**
       Shrink the angle $\alpha$ as: **If** $\alpha < 0$; $\alpha_{\min} = \alpha$; **Else** $\alpha_{\max} = \alpha$;
       Re-draw $\alpha \sim \text{Unif}[\alpha_{\min}, \alpha_{\max}]$;
       **GoTo** (4);
   **Endif**

   **Endfor**

   **Endfor**

---

### 3.2.3 Termination condition

In the SeMIS algorithm, $I$ proposal distributions are involved for the evidence estimation and posterior sampling. While the choice of $I$ is not critical to SeMIS, selecting an appropriate $I$ can help to improve the performance of the overall algorithm. In this paper, the value of $I$ is chosen such that final parameter $\lambda_I$ approaches 1, ensuring the last proposal distribution $q_{I-1}(\boldsymbol{\theta};\lambda_{I-1})$ closely approximate the posterior distribution $p(\boldsymbol{\theta}|\boldsymbol{D})$. Practically, iteration stops when $\ln r_{I-1} \geq -10^{-4}$ (i.e., $r_{I-1} = \exp(-10^{-4}) \approx 1$), a condition guaranteed by the monotonic increase of $r_i$ as $i$ becomes large. Specifically, as $i$ grows, $l_{0:i-2}^{\max} \approx l_{0:i-1}^{\max}$ ensures $r_i > r_{i-1} l_{0:i-2}^{\max}/l_{0:i-1}^{\max} > r_{i-1}$.

A direct consequence of this termination is that proposal distributions $q_0(\boldsymbol{\theta};\lambda_0)$, $q_1(\boldsymbol{\theta};\lambda_1)$, ..., $q_{I-1}(\boldsymbol{\theta};\lambda_{I-1})$ define a smooth transition from the prior to the posterior distribution. Since $q_{I-1}(\boldsymbol{\theta};\lambda_{I-1})$ effectively becomes the posterior distribution, one can find an equivalent expression of evidence by comparing expressions of $q_{I-1}(\boldsymbol{\theta};\lambda_{I-1})$ and $p(\boldsymbol{\theta}|\boldsymbol{D})$ as

$$z = \wp_{I-1} L_{\sup} \tag{23}$$

Based on the principle of importance sampling, one can find

$$\begin{aligned}
\wp_i &= \int \min\left[\frac{L(\boldsymbol{\theta})}{\lambda_i L_{\sup}}, 1\right] \pi(\boldsymbol{\theta}) \mathrm{d}\boldsymbol{\theta} \\
&= \int \frac{\min\left[\frac{L(\boldsymbol{\theta})}{\lambda_i L_{\sup}}, 1\right]}{\wp_{i-1}^{-1} \min\left[\frac{L(\boldsymbol{\theta})}{\lambda_{i-1} L_{\sup}}, 1\right]} \wp_{i-1}^{-1} \min\left[\frac{L(\boldsymbol{\theta})}{\lambda_{i-1} L_{\sup}}, 1\right] \pi(\boldsymbol{\theta}) \mathrm{d}\boldsymbol{\theta} \\
&= \wp_{i-1} \mathbb{E}_{q_{i-1}(\boldsymbol{\theta};\lambda_{i-1})}[\beta_i(\boldsymbol{\theta})] \\
&= \wp_{i-1} p
\end{aligned} \tag{24}$$

where we have used Eqn. (22) and the setting $\mathbb{E}_{q_{i-1}(\boldsymbol{\theta};\lambda_{i-1})}[\beta_i(\boldsymbol{\theta})] = p$. This gives a recursive relation $\wp_{I-1} = p^{I-1}$ (with $\wp_0 = 1$), leading to $z = p^{I-1} L_{\sup}$. In practice, we do not know $L_{\sup}$ and the exact value $p$, and thus need to estimate them from random samples in SeMIS $\{\boldsymbol{\Theta}_{i,k}, i = 0,1,...,I-1,\ k = 1,2,...,N_i\}$. In terms of log evidence, we propose the following estimator

$$\ln \hat{Z}_{\text{SIS}} = \ln \hat{L}_{0:I-1}^{\max} + \sum_{i=0}^{I-1} \ln \left[ N_i^{-1} \sum_{k=1}^{N_i} \beta_i(\boldsymbol{\Theta}_{i,k}) \right] \quad (25)$$

where $\hat{L}_{0:I-1}^{\max}$ is an estimator of $L_{\sup}$. Since the above estimator follows the principle of sequential importance sampling (SIS) [32], we denote it as the SIS estimator for the evidence, and report it alongside the MIS estimator for cross-validation.

## 4. Empirical Studies

In this section, we investigate the performance of the proposed SeMIS algorithm for Bayesian inference. We first compare SeMIS with two state-of-the-art approaches i.e., SuS [14] and adaptive BUS (aBUS) [17], in terms of evidence estimation and posterior sampling. We then consider an example on finite element (FE) model updating to illustrate the performance of SeMIS for a practical high-dimensional and possibly multimodal problem.

**4.1 Three benchmark examples**

Three benchmark Bayesian inference problems characterized by high dimensionality and multimodality are first considered, as described in Table 1 and illustrated in Figure 2 (for the case of Dimension 2). Example A (Eggbox) contains a large number of modes, making it challenging to generate samples. In Example B (Gaussian shells), the posterior probability mass is concentrated within two thin "shells", which can be interpreted as an infinite number of connected modes. Example C (Normal-LogGamma mixture) has four widely separated modes, with an analytical solution available. In terms of identifiability, both Examples A and C are locally identifiable, while Example B is unidentifiable.

We run SuS, aBUS, and SeMIS with the same level of probability $p = 0.1$, and repeat each algorithm for 1,000 times. To ensure a fair comparison between three algorithms, we control the number of likelihood function evaluations ($N_{cal}$) to be approximately equal. Specifically, we set a sample size $N = 500$ for SuS, and the sample sizes of aBUS and SeMIS are accordingly determined to achieve a similar $N_{cal}$.

Table 1. Three benchmark problems

| Example | Dimension | Likelihood function | Prior Distribution |
|---|---|---|---|
| A. Eggbox [33] | 2 | $L(\boldsymbol{\theta}) = \exp\left[2 + \cos\left(\frac{\theta_1}{2}\right)\cos\left(\frac{\theta_2}{2}\right)\right]^5$ | Unif$(0, 10\pi)$ |
| B. Gaussian shells [34] | 2, 10, 30 | $L(\boldsymbol{\theta}) = \text{circ}(\boldsymbol{\theta}; \boldsymbol{c}_1, r_1, w_1) + \text{circ}(\boldsymbol{\theta}; \boldsymbol{c}_2, r_2, w_2)$ <br> $\text{circ}(\boldsymbol{\theta}; \boldsymbol{c}, r, w) = \frac{1}{\sqrt{2\pi w^2}}\exp\left[-\frac{(|\boldsymbol{\theta}-\boldsymbol{c}|-r)^2}{2w^2}\right]$ <br> $\boldsymbol{c}_1 = [-3.5, \mathbf{0}_{1\times n-1}],\ \boldsymbol{c}_2 = [3.5, \mathbf{0}_{1\times n-1}]$ <br> $w_1 = w_2 = 0.1,\ r_1 = r_2 = 2$ | Unif$(-6, 6)$ |
| C. Normal-LogGamma mixture [34] | 2, 5, 10, 20 | $L(\boldsymbol{\theta}) = \prod_{i=1}^{n} L(\theta_i)$ <br> $L(\theta_1) = 0.5LG(\theta_1|-10,1,1) + 0.5LG(\theta_1|10,1,1)$ <br> $L(\theta_2) = 0.5N(\theta_2|-10,1) + 0.5N(\theta_2|10,1)$ <br> For $3 \leq i \leq (n+2)/2,\ L(\theta_i) = LG(\theta_i|10,1,1)$ <br> For $(n+2)/2 \leq i \leq n,\ L(\theta_i) = N(\theta_i|10,1)$ | Unif$(-30, 30)$ |

Note: "LG", "N" and "Unif" denote the log Gamma, normal, and uniform distribution, respectively.

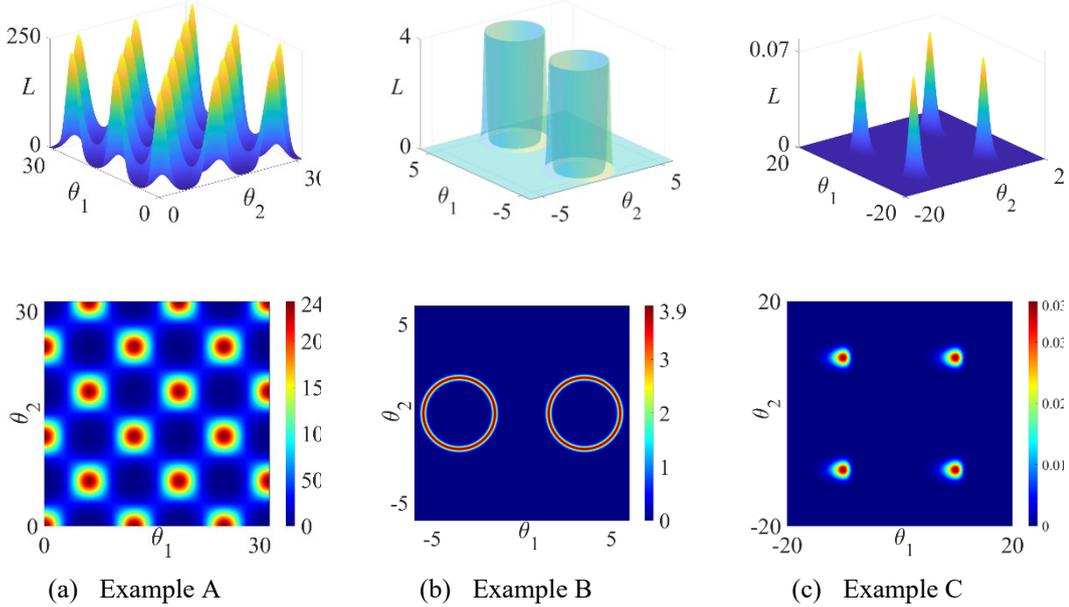

(a) Example A    (b) Example B    (c) Example C

Figure 3. 3-D shape and contour plots of likelihood functions; three benchmark examples [14].

The evidence estimations of three algorithms are listed Table 2 in terms of the relative bias of sample mean and coefficient of variation (c.o.v.). Two evidence estimates (i.e., MIS and SIS) from SeMIS are also reported for crosscheck. Values in the second and third columns represent the dimension of considered problems and the analytical log evidence, respectively. It is seen that the relative biases of all algorithms are less than 1%, showing good performance of them. Of three algorithms, the two estimates from SeMIS yields the smallest bias and c.o.v. in most examples, validating the strategy in SeMIS to improve its precursor SuS. As for the two estimators from SeMIS, they show similar bias values, but the c.o.v. of MIS is smaller than that of SIS in most cases.

Table 2. Evidence estimation results of SuS, aBUS and SeMIS; three benchmark examples.

| Example | Dimension | Reference log evidence | Relative bias of mean estimate [‰] (c.o.v. [%]) | | | |
| --- | --- | --- | --- | --- | --- | --- |
| | | | SuS | aBUS | SeMIS | |
| | | | | | MIS | SIS |
| A | 2 | 235.86 | -0.08 (0.11) | -0.04 (0.13) | -0.08 (0.09) | -0.05 (0.09) |
| B | 2 | -1.75 | -1.03 (7.92) | 5.71 (14.2) | -2.59 (7.44) | -3.17 (7.41) |
| | 10 | -14.59 | 0.84 (2.90) | 0.69 (3.42) | -2.93 (2.67) | -5.58 (3.42) |
| | 30 | -60.13 | 1.66 (1.48) | 1.66 (1.78) | 0.64 (1.46) | -0.23 (1.47) |
| C | 2 | -8.19 | 2.44 (1.95) | 2.44 (3.05) | 2.10 (1.29) | 1.84 (1.36) |
| | 5 | -20.47 | -0.49 (1.71) | 1.95 (2.44) | 1.03 (1.27) | 0.97 (1.33) |
| | 10 | -40.94 | 3.66 (1.63) | 3.42 (2.34) | 1.62 (1.13) | 1.02 (1.13) |
| | 20 | -81.89 | 6.35 (1.66) | 7.45 (1.83) | 5.81 (1.58) | 5.07 (1.59) |

Note: Relative bias of mean estimate is calculated as sample mean/reference log-evidence - 1, and the value in parenthesis shows the sample c.o.v. in percentage.

For posterior sampling, we choose Example C to plot posterior samples and compare with analytical posterior marginal distributions, as shown in Figure 4. The dimension of Example C is set to 20, and six dimensions (1, 2, 3, 11, 12, and 20) are selected for visualization. Since SuS demonstrates a superior performance in posterior sampling compared with aBUS [14], we focus here on the comparison of SeMIS and SuS. For a better visualization, a larger sample size is considered while maintaining similar computational cost ($N_{cal} = 1.21 \times 10^6$ in SuS and $N_{cal} = 1.19 \times 10^6$ in SeMIS). Here, the red curve in the histogram plot represents the analytical marginal PDF, and the heatmap represents the sample density of any two-dimensional joint distribution. In the first and second dimensions, both algorithms effectively recover the mixed LogGamma and mixed Gaussian distributions, respectively. Artificial spikes may appear due to repeated samples in the importance resampling step.

Since it is difficult to compare the performance of SeMIS and SuS visually, we further employ the Kolmogorov-Smirnov (K-S) statistic to quantify the difference between the simulated posterior marginal distributions and the analytical ones. The K-S statistic computes the largest difference between the empirical cumulative distribution function (CDF) and the analytical CDF, thus a smaller value of K-S statistic indicates a better matching between them. The results are presented in Table 3, showing that the K-S statistic of SeMIS is generally smaller than that of SuS and aBUS. The advantage of SeMIS is particularly evident in Dimensions 1 and 2, which contain mixed LogGamma and mixed Gaussian distributions, respectively. This demonstrates that SeMIS is more effective in multimodal posterior sampling, because its proposal distributions allow an easier transition between different modes.

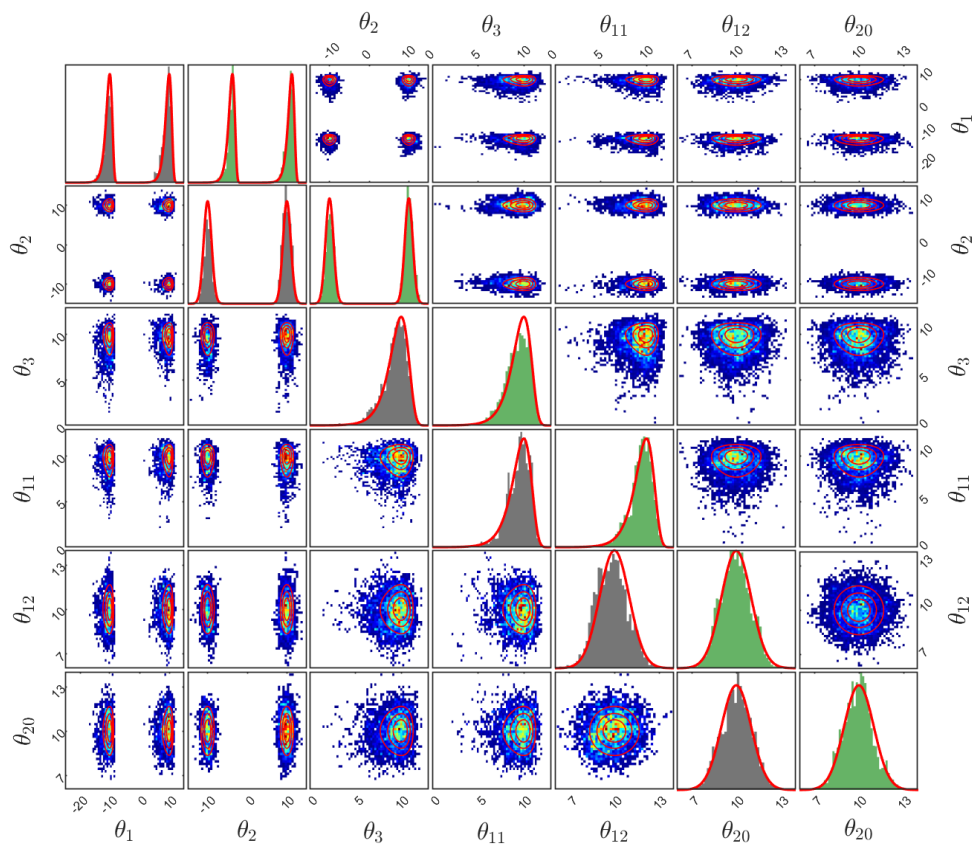

Figure 4. Posterior simulation of SuS and SeMIS; Example C

Note: Lower-left and upper-right part show the result of SuS and SeMIS, respectively; Gray and green histogram represent the approximated marginal distributions given by SuS and SeMIS, respectively; Red solid curve indicates the analytical posterior marginal distributions.

Table 3. K-S statistics in Example C.

| Dimension | SuS | aBUS | SeMIS |
|---|---|---|---|
| 1 | 0.1874 | 0.1884 | 0.1628 |
| 2 | 0.2068 | 0.1863 | 0.1618 |
| 3 | 0.0554 | 0.0901 | 0.0573 |
| 11 | 0.0590 | 0.0916 | 0.0546 |
| 12 | 0.0520 | 0.0850 | 0.0461 |
| 20 | 0.0530 | 0.0862 | 0.0488 |

To quantify the sampling efficiency of all three methods, we calculate the ESS ratio $N_{ess}/N_{cal}$, i.e., the ESS per evaluation of likelihood function. The ESS $N_{ess}$ is calculated based on Eqn. (9) for SeMIS with the assumption of statistical independence between samples. For a fair comparison, $N_{ess}$ of SuS and aBUS is also calculated by assuming sample independence [14]. The average value of ESS ratio $N_{ess}/N_{cal}$ is presented in Table 4 for 1,000 repeated computations. It is seen that SeMIS performs better than SuS and aBUS in Examples A and C across all dimensions, while SuS yields the best result in Example B. One can understand the discrepancies in performance by the geometry of the likelihood function. In Examples A and C, all modes are isolated by low-likelihood regions, whereas, in Example B, all modes are directly connected, allowing an easy transition between modes. This diminishes the effectiveness of SeMIS, making it slightly less efficient compared to SuS in this particular case. For the value of ESS $N_{ess}$, SeMIS yields no less than 900 for Examples A and C, which is much larger than SuS ($\geq 600$) and aBUS ($\geq 300$). However, all methods perform suboptimally in Example B, where $N_{ess} \leq 400$. A prior study [35] suggests that an ESS below 400 may compromise estimation reliability. Notably, the ESS values reported here represent upper bounds; thus, the actual performance may be even less favorable. To address this limitation in SeMIS, an additional MCMC sampling step can be applied directly to $q_{I-1}(\boldsymbol{\theta}; \lambda_{I-1})$ - an approximated posterior with negligible difference. Since seeds are already distributed as $q_{I-1}(\boldsymbol{\theta}; \lambda_{I-1})$, no burn-in phase is required, improving efficiency.

Table 4. Average value of ESS ratio $N_{ess}/N_{cal}$; three benchmark examples.

| Example | Dimension | SuS [%] | aBUS [%] | SeMIS [%] |
|---|---|---|---|---|
| A | 2 | 3.70 (15.4) | 2.06 (15.1) | 5.82 (15.6) |
| B | 2 | 12.6 (2.67) | 5.50 (2.74) | 12.8 (2.64) |
|  | 10 | 1.90 (23.7) | 1.52 (22.5) | 1.44 (20.5) |
|  | 30 | 0.54 (126) | 0.34 (111) | 0.38 (113) |
| C | 2 | 5.79 (10.4) | 3.07 (10.9) | 11.5 (11.6) |
|  | 5 | 2.31 (37.8) | 0.84 (38.3) | 4.70 (38.2) |
|  | 10 | 1.27 (96.0) | 0.34 (98.8) | 2.81 (98.5) |
|  | 20 | 0.75 (227) | 0.16 (229) | 1.24 (227) |

Note: the value in parentheses indicates the average number of likelihood evaluation $N_{cal}$ [$\times 10^3$].

As seen in Table 4, the ESS ratio $N_{ess}/N_{cal}$ decreases with increasing dimension, reflecting the greater computational effort needed to generate posterior samples - a manifestation of the curse of dimensionality in Bayesian inference. This behavior can be explained by examining the number of likelihood evaluations $N_{cal}$, which scales linearly with dimension (exemplified by Example C). The linear relationship arises because the distance between the prior and posterior distributions grows exponentially with dimension [36], while the number of required proposal distributions $I$ increases only logarithmically with this distance (since $p_0 = 1$ for prior and $p_{I-1} = p^{I-1}$ for posterior). Together, these combined trends result in the observed linear dependence of $N_{cal}$ on dimension.

**4.2 Finite element model updating**

This section presents an application of the proposed SeMIS algorithm for finite element model (FEM) updating of the IASC-ASCE SHM benchmark model [37]. The considered structure is a quarter-scale model of a four-story, two-bay by two-bay steel-frame building, measuring 3.6 m in height and 2.5 m by 2.5 m in plan. A schematic diagram of the model is provided in Figure 5(a). A 120 degrees-of-freedom (DoFs) FEM [37] is employed to simulate the structural response data. Independent Gaussian white noise excitation is applied at all DOFs. Modal superposition method is used to calculated the dynamic responses with a damping ratio of 1% for all modes. The dataset comprises acceleration measurements from 16 channels, situated at the midpoint of each story's four edges as depicted in Figure 5(a). The dataset spans a total duration of 100 seconds, with a sampling frequency of 200 Hz.

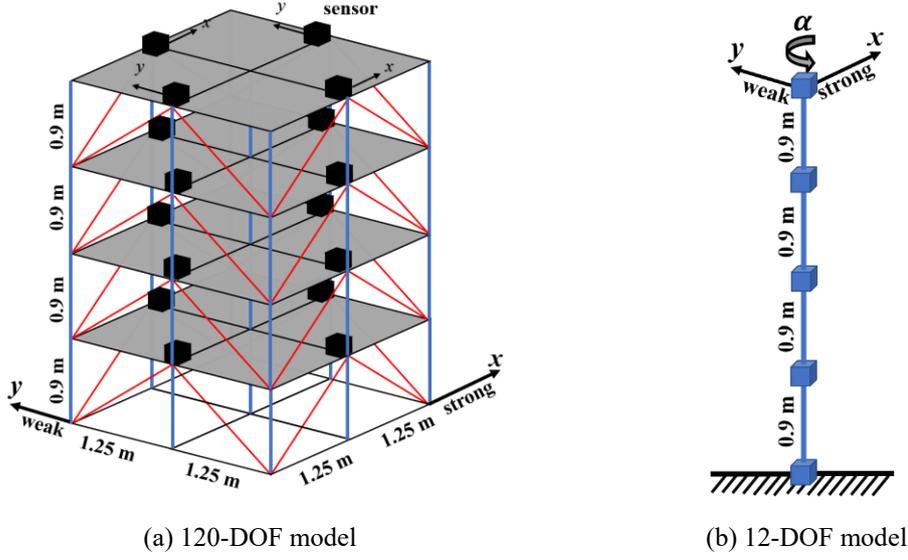

(a) 120-DOF model          (b) 12-DOF model

Figure 5. Schematic diagram of considered building model [37]

A reduced order model, consisting of two translational DoFs and one rotational DoF for each floor, is considered to update, as depicted in Figure 5(b). The initial values of inter-story stiffness ($k_{x,i}$, $k_{y,i}$, and $k_{\alpha,i}$ for $i = 1,2,3,4$) and floor mass ($m_i$ and $I_i$ for $i = 1,2,3,4$) are obtained from Ref. [37]. The unknown parameter is represented as a multiplier of inter-story stiffness or floor mass, e.g., the updated inter-story stiffness $k'_{x,i} = \theta_{x,i} k_{x,i}$ and floor mass $m'_i = \theta_{m,i} m_i$. All 20 unknown parameters are organized into a vector $\boldsymbol{\theta} = [\theta_{x,1}, \theta_{y,1}, \ldots, \theta_{\alpha,4}, \theta_{m,1}, \ldots, \theta_{I,4}]^{\mathrm{T}}$ for simplicity.

A two-stage scheme is adopted for FEM updating, where the FEM parameter $\boldsymbol{\theta}$ is updated based on identified modal parameters (frequencies $f_m$ and mode shapes $\boldsymbol{\varphi}_m$) from "measured" structural accelerations in the first stage. To be consistent, a Bayesian modal identification approach [38,39] is applied to obtain a normal distribution approximation of the posterior distribution of modal parameters, with the mean vector $\hat{\boldsymbol{\omega}} = [\hat{f}_1, \hat{\boldsymbol{\varphi}}_1^{\mathrm{T}}, \ldots, \hat{f}_M, \hat{\boldsymbol{\varphi}}_M^{\mathrm{T}}]^{\mathrm{T}}$ and the covariance matrix $\hat{\boldsymbol{C}}$. In the second stage, the identified modal parameters $\hat{\boldsymbol{\omega}}$ are regarded as "data", and the following log likelihood function is applied

$$\mathcal{L}(\boldsymbol{\theta}; \hat{\boldsymbol{\omega}}) = \ln L(\boldsymbol{\theta}; \hat{\boldsymbol{\omega}}) = -[\hat{\boldsymbol{\omega}} - \boldsymbol{\omega}(\boldsymbol{\theta})]^{\mathrm{T}} \hat{\boldsymbol{C}}^{-1} [\hat{\boldsymbol{\omega}} - \boldsymbol{\omega}(\boldsymbol{\theta})]/2 \qquad (26)$$

where $\boldsymbol{\omega}(\boldsymbol{\theta})$ denotes the modal parameters calculated from the FEM for a specific value of $\boldsymbol{\theta}$ via eigenvalue decomposition. In this paper, we assume an independent uniform prior Unif(0,1.5) for each stiffness parameter and Unif(0.9,1.1) for each mass parameter, reflecting different confidence on stiffness and mass parameters.

We first consider a full setup with all measurement points and the first twelve modes (i.e., $M = 12$) for FEM updating. We apply SeMIS, SuS and aBUS algorithms for identification, with the same setting as in Section 4.1. That is, we set a sample size $N = 500$ for SuS, and the sample sizes of aBUS and SeMIS are accordingly determined to achieve a similar $N_{cal}$ for a fair comparison. The identified results are shown in Figure 6 in terms of the posterior mean and three standard deviations. It is seen that all algorithms yield similar results, cross-validating each other. The calculated log evidences are also similar with $-1.558 \times 10^3$, $-1.565 \times 10^3$ and $-1.573 \times 10^3$, respectively, for SeMIS, SuS and aBUS. Mass parameters are considered as updating parameters, while they are considered to be known in the original study [37]. To validate this setting, we also consider to update only the stiffness parameters, yielding the log evidences as $-2.185 \times 10^3$, $-2.184 \times 10^3$ and $-2.187 \times 10^3$ for SeMIS, SuS and aBUS, respectively. Since the evidence of the model with both stiffness and mass updated is significantly larger, updating the mass parameter is beneficial following the principle of Bayesian model selection [38].

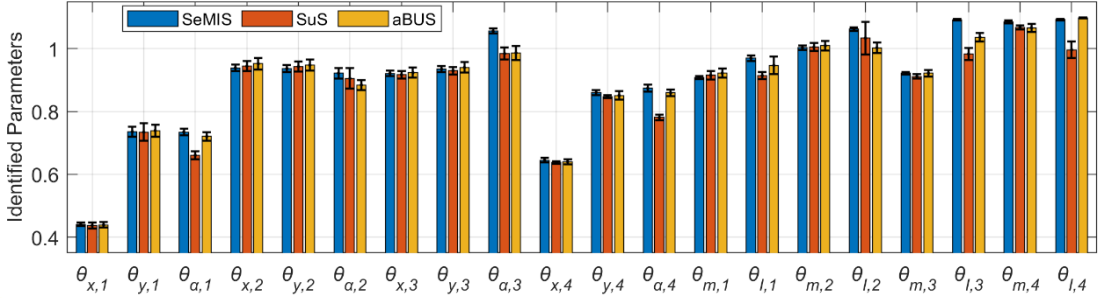

Figure 6. Identified FEM parameters.
Note: height of each bar represents the posterior mean, and error bar shows ±3 standard deviations.

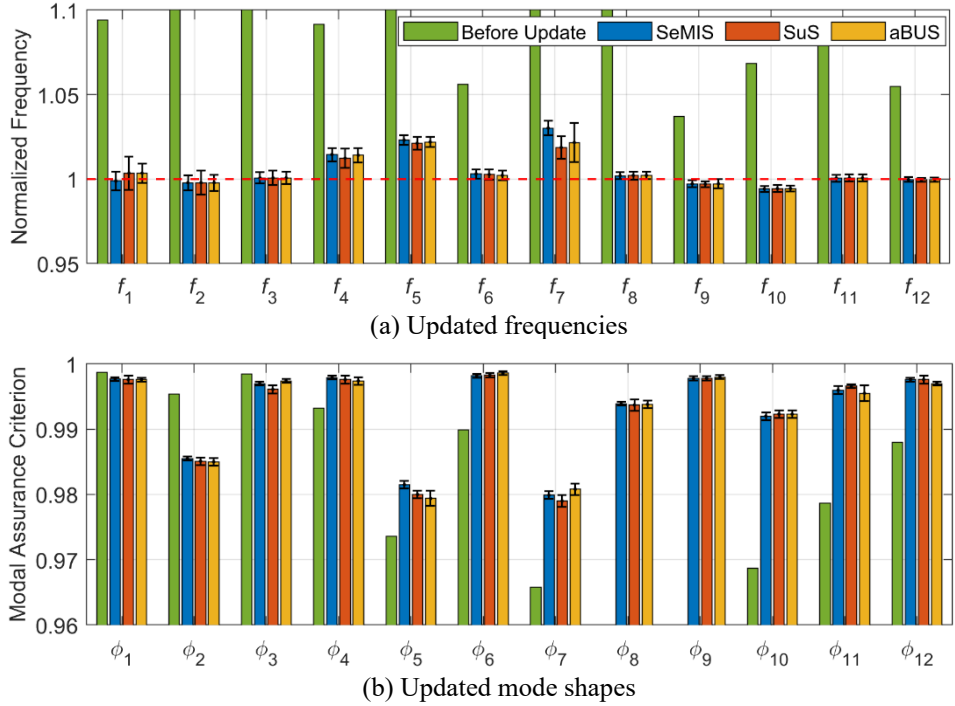

Figure 7. Performance of updated FEM.

The performance of updated FEM is illustrated in Figure 7 in terms of updated frequencies and mode shapes. The normalized frequency represents the frequency ratio of the 12-DoF FEM and the identified value from data. Although the initial difference is large in frequencies, all updated frequency values are close to the actual ones, with relative errors below 3%. The modal assurance criterion (MAC) denotes the cosine angle between the mode shapes calculated from the 12-DoF FEM and identified from data. The first three MAC values slightly decrease after update, but the remaining ones have been increased to be no less than 0.98, especially for Modes 8 and 9. The consistence between updated and measured modal parameters validates the performance of the proposed SeMIS algorithm for Bayesian inference.

Besides the original FEM (denoted as Pattern 0), we also consider two damage cases: removing the first-story braces (Pattern 1) and removing the first- and third-story braces (Pattern 2), and try to identify those damages by comparing with the updated FEMs. In addition, we consider a scenario where only partial DoFs and modes are measured. In this case, sensors on the second and fourth stories are used, and modal data of the first six modes are selected for Bayesian inference. The posterior samples of $\boldsymbol{\theta}$ ($\theta_{x,3}$ and $\theta_{y,3}$) for all considered cases are presented in Figure 8, showing the reduction of stiffness parameters due to damages. When no damage is present, the stiffness parameters are close to 1 and decrease to different extents depending on the damage pattern. When sufficient data is available, the posterior sample distribution follows a unimodal pattern. However, with only partial data, the posterior distribution differs significantly from that of sufficient data and may exhibit multimodality, indicating unreliable results.

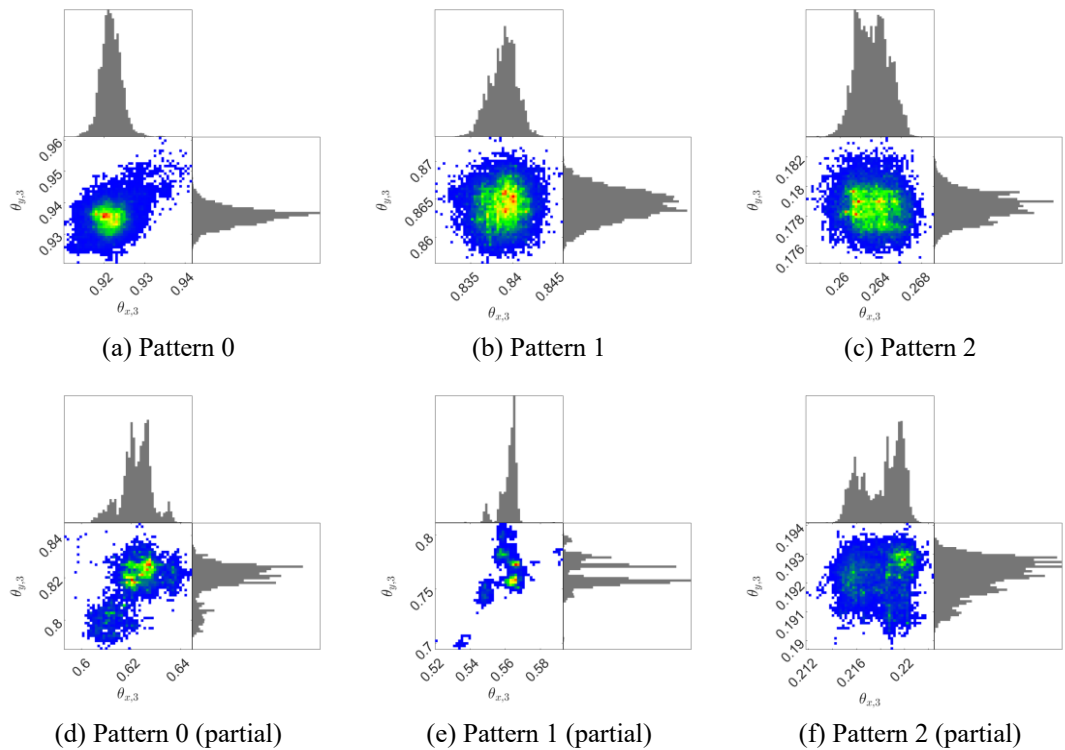

(a) Pattern 0    (b) Pattern 1    (c) Pattern 2

(d) Pattern 0 (partial)    (e) Pattern 1 (partial)    (f) Pattern 2 (partial)

Figure 8. Posterior samples of three damage patterns.

Based on identified stiffness parameters, we conducted an analysis on damage detection and localization. For this, we consider stiffness parameters of Pattern 0 as the reference, and calculate the stiffness change in Patterns 1 and 2, shown in Figure 9. All algorithms convey evident stiffness loss (surpassing 40%) in floors where damage is present in Patterns 1 and 2. Notably, this stiffness loss is in close agreement with the theoretical values. Conversely, the stiffness at undamaged locations only shows minor fluctuations, all measuring below 15%. This indicates that updating the FEM parameters with the proposed SeMIS algorithm is applicable for damage detection and localization. Given their shared foundation in Bayesian inference and utilization of consistent datasets, the SeMIS, SuS, and aBUS algorithms demonstrate comparable performance characteristics, cross-validating each other.

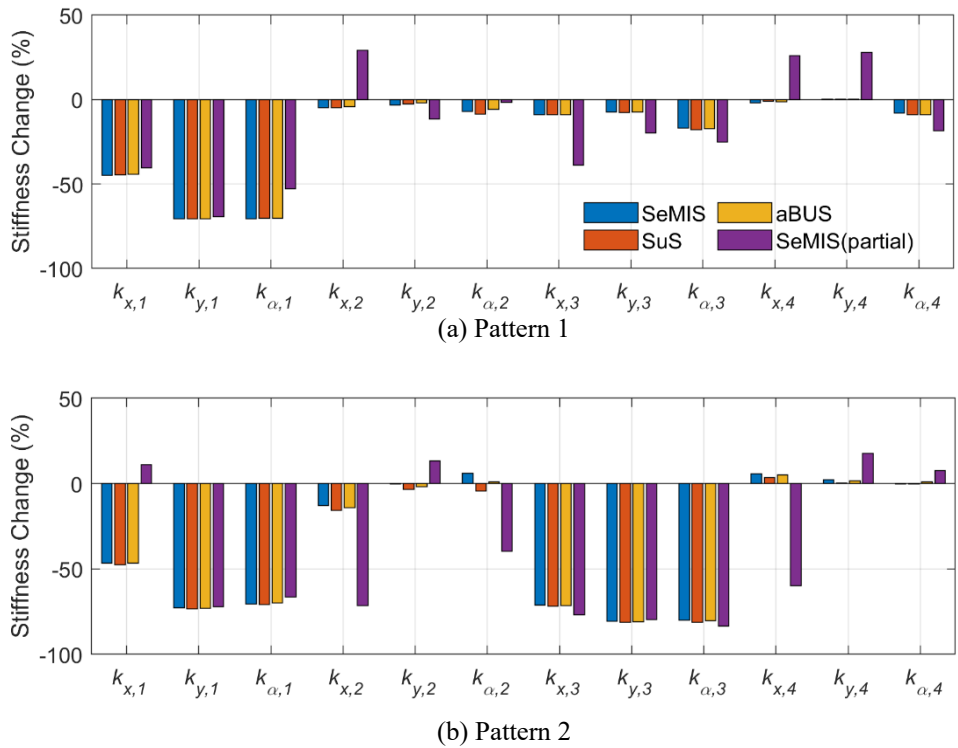

Figure 9. Damage detection based on stiffness change.

Finally, misleading results in evaluating the stiffness change may be possible when only partial data is used in FEM updating. For example, a notable stiffness reduction of more than 50% is present in the second and fourth stories in Pattern 2. Such pronounced stiffness reduction in undamaged stories could potentially lead to erroneous damage assessments. Using only limited sensors and modal data for damage localization process may result in false alarms.

## 5. Conclusions

This paper presents the sequential multiple importance sampling (SeMIS) algorithm for high-dimensional Bayesian inference, integrating softly truncated prior distributions and balance heuristic weighting within the multiple importance sampling (MIS) framework. SeMIS enables adaptive proposal distribution tuning and hyperparameter determination, effectively facilitating sample transitions between high- and low-likelihood regions-particularly crucial for multimodal posterior exploration.

Through rigorous benchmarking against subset simulation (SuS) and adaptive Bayesian updating with structural reliability (aBUS), SeMIS demonstrates superior performance in both evidence estimation and posterior sampling. Specifically, it achieves lower bias and coefficient of variation (c.o.v.) in evidence computation, while posterior samples exhibit closer alignment with the true distribution (lower Kolmogorov-Smirnov statistic) and higher efficiency (larger effective sample size ratio). These advantages are especially pronounced for multimodal problems with well-separated modes. The practical utility of SeMIS is underscored by its successful application to a high-dimensional finite element model (FEM) updating problem in civil engineering. The SeMIS algorithm accurately localizes structural damage by quantifying stiffness loss, highlighting it as a robust tool for uncertainty quantification in structural health monitoring.

By bridging methodological advancements in Bayesian computation with critical engineering applications, SeMIS offers a scalable and efficient framework for tackling high-dimensional inference challenges. However, two key theoretical aspects require further investigation: (1) derivation of the analytical formula for estimation uncertainty in evidence computation, and (2) development of a more precise expression for effective sample size. Future work will address these limitations while extending applications to real-time structural monitoring systems and other large-scale inverse problems in civil infrastructure.

## CRediT authorship contribution statement

**Binbin Li**: Writing - review & editing, Writing - original draft, Visualization, Validation, Supervision, Software, Resources, Project administration, Methodology, Investigation, Funding acquisition, Formal analysis, Data curation, Conceptualization.

**Xiao He**: Writing - review & editing, Writing - original draft, Visualization,

Validation, Software, Methodology, Investigation, Formal analysis, Data curation.

**Zihan Liao**: Writing - review & editing, Visualization, Validation, Software, Methodology, Investigation, Formal analysis, Data curation, Conceptualization.

# Declaration of competing interest

The authors declare that they have no known competing financial interests or personal relationships that could have appeared to influence the work reported in this paper.

# Data availability

The data and code will be made available once the paper is accepted for publication.

# Acknowledgement

This work is partially supported by the National Natural Science Foundation of China (U23A20662). Any opinions, findings, conclusions, or recommendations expressed in this material are those of the authors and do not reflect the views of the funder.